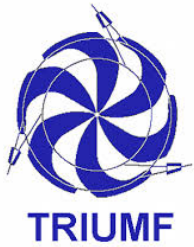
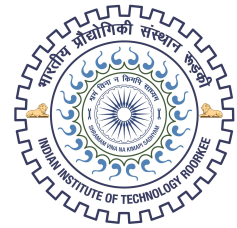

INDIAN INSTITUTE OF TECHNOLOGY ROORKEE
ROORKEE, UTTARAKHAND, INDIA 247667

MASTERS DISSERTATION

---

# γ-ray Angular Distributions in Single Nucleon Transfer Reactions with Exotic Strontium Isotopes

---

*Author:*
Rahul Sunil Jain

*Supervisor:*
Prof. Reiner Krücken
*Co-Supervisor*:
Prof. Moumita Maiti

*A dissertation submitted in partial fulfillment of the requirements*
*for the degree of*
Integrated Masters of Science
*in the*
Subject of Physics
*to the*
Department of Physics
Indian Institute of Technology Roorkee

# Declaration of Authorship

I, Rahul Sunil Jain, declare that this dissertation titled, "$\gamma$-ray Angular Distributions in Single Nucleon Transfer Reactions with Exotic Strontium Isotopes" in partial fulfillment of the requirements for the award of the degree of Integrated Masters of Science in Physics to the Department of Physics, Indian Institute of Technology Roorkee is an authentic record of my original research work carried out during the period of January - April 2018 under the supervision of Prof. Reiner Kruecken at TRIUMF Laboratory, Vancouver, Canada and the co-supervision of Prof. Moumita Maiti, Department of Physics, Indian Institute of Technology Roorkee.

I confirm that this work has not been submitted by me to any other university/institute for the award of any academic degree.

Signed:
___________________________________

Rahul Sunil Jain
Integrated MSc. Physics (Final year)
Department of Physics
Indian Institute of Technology Roorkee

Date:
___________________________________

# Certificate

This is to certify that the dissertation titled, "$\gamma$-ray Angular Distributions in Single Nucleon Transfer Reactions with Exotic Strontium Isotopes" prepared by Mr. Rahul Sunil Jain is a record of his own original research work carried out under our supervision and guidance. This work is being presented to the Department of Physics, Indian Institute of Technology Roorkee in partial fulfillment for the award of the Integrated Masters of Science in Physics degree.

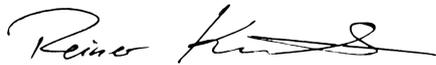

Prof. Reiner Krücken
Supervisor
Department of Physics and Astronomy
University of British Columbia, Vancouver

Prof. Moumita Maiti
Co-Supervisor
Department of Physics
Indian Institute of Technology Roorkee

## *Abstract*

$\gamma$-ray angular distributions help assign spin and parity to excited energy levels in nuclei. The spectroscopy of $^{94,96}$Sr (Cruz 2017) studied through the single neutron transfer reactions with $^{95}$Sr beam in inverse kinematics [$^{95}$Sr(d,p)$^{96}$Sr; $^{95}$Sr(d,t)$^{94}$Sr] revealed a rich nuclear structure with many excited states. While the spin-parities were assigned to low lying states through techniques like particle angular distributions and angular momentum transfer, those for higher lying states were ambiguous. The goal of this project is to develop the $\gamma$-ray angular distributions and correlations techniques to assign spin-parity to these states.

In this work, the $\gamma$-ray angular distributions for 815 keV transition from the first excited state to ground state ($2_1^+ \rightarrow 0^+$) in $^{96}$Sr are measured. The alignment of this state is calculated from these measurements and compared to the theoretically calculated alignment for this state which is determined by coupling the spins of reactants and the orbital angular momentum transferred. The values agree within experimental limits justifying the technique. It is then applied to some other transitions and important results are discussed. Angular correlations cannot be performed with this data since statistics turn out to be limiting factor at the crystal level.

# *Acknowledgements*

This dissertation would never have been completed without the contributions of a large number of people who helped me throughout this work and before that. I would try my best to acknowledge all of them.

First of all, I would like to express my sincere gratitude towards my supervisor Prof. Reiner Krücken for providing me the opportunity to work at a world-class facility like TRIUMF. I would thank Dr. Greg Hackman for the very useful discussions and suggestions provided by him which kept the project going in the right direction. I would also like to thank Dr. Soumendu Bhattacharjee for supervising my research and providing valuable inputs as and when required.

I would like to thank the $\gamma$-ray spectroscopy group at TRIUMF to host and support me during Spring 2018 semester and providing me a conducive environment for research. Also many thanks to the entire GAPS committee - graduate students and Post-docs, who were always ready to help at and outside of work. I am also thankful to my roommates and amazing friends from Vancouver who made my time in Canada a memorable one.

My sincere thanks to the Department of Physics at the Indian Institute of Technology Roorkee, who shaped and kept my interest in Physics thriving for the last five years. I am grateful to my Co-Supervisor, Prof. Moumita Maiti for supporting me in my endeavors. I am also thankful to the faculty at IIT Roorkee who taught and guided me throughout my undergraduate and masters studies. I am deeply grateful to my wonderful friends from IIT Roorkee, who have been there with me for last five years. Special thanks to Vatsal for encouraging and helping me to learn and use LaTeX to write this document.

I would like to thank my family, who have supported me throughout, in whatever I wanted to do and always kept my motivation high whenever I felt low.

# Contents

# List of Figures

*Dedicated To my Parents who continue to support me at every stage in my life.*

# Chapter 1

# Introduction

## 1.1   Motivation

Nuclear shell model is by far the most successful in describing the atomic nucleus. It can accurately predict the magic numbers, ground state spin-parity of even as well as odd mass number nuclei, etc. Moreover, the shell structure of nuclei plays an important role in the propensity for nuclei to deform. Even a small number of valence nuclei outside of a closed shell can drive the whole nucleus into a deformed shape. This is evident from Figure 1.1 which shows the predicted ground state quadrupole deformation across the nuclear chart (Moller et al. 1995). The region around Z 40 and N 60 has protons as well as neutrons outside the closed shell and hence, high deformation.

Another interesting feature of this region in the nuclear chart is the shape co-existing $0^+$ states. These states are quantum superposition of spherical and deformed $0^+$ states which can interact strongly with each other. An experimental signature of this phenomenon is the enhanced monopole (E0) transition strengths which can proceed only through internal conversion electrons mechanism. The shape-coexisting states in these nuclei can be analyzed through a very simple yet accurate two-level mixing model. It assumes that ground states of spherical and deformed nuclei initially exist with a separation $\Delta E_u$. An interaction $V$ acts between them in a way that pushes the two levels further apart with final separation $\Delta E_p$. Their wavefunctions mix with the mixing amplitude $a$ such that

$$|0_2^+> = a|0_{sph}^+> + \sqrt{1-a^2}|0^{+0_{def}}>$$
$$|0_3^+> = \sqrt{1-a^2}|0_{sph}^+> - a|0^{+0_{def}}> \tag{1.1}$$

$$\Delta E_p = \sqrt{(\Delta E_u)^2 + 4V^2} \tag{1.2}$$

A large number of theoretical calculations exist for this region with various models predicting different results. However, this part of the nuclear chart remained experimentally unexplored. With the advancement in rare isotope beam production capabilities, the experimental results from the spectroscopy of this region of nuclear chart is indispensable to validate the theoretical models.

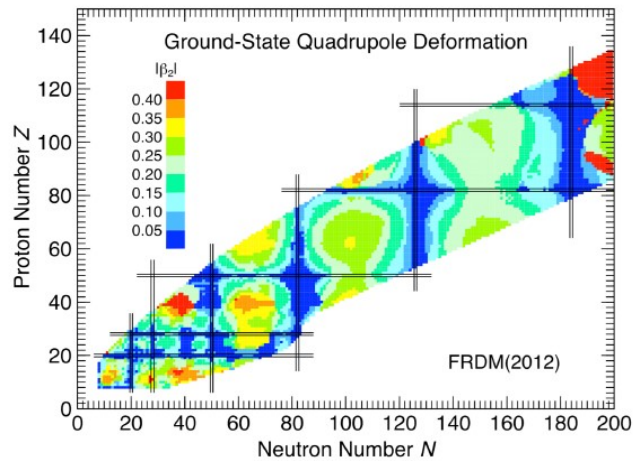

FIGURE 1.1: Finite Range Drop Model (FRDM) calculations of the ground-state deformation across the nuclear chart by Moller et al. (1995).

## 1.2 Transfer Reactions

Transfer reactions are a well-established tool for studying single particle structure of nuclei. In these reactions, a small number of valence nucleons are transferred from either the target to the projectile or vice-versa. This leads to two types of transfer reactions: stripping and pickup. Stripping reactions are when some nucleons are stripped by the projectile from the target whereas pickup is when some nucleons are picked up by the projectile from target nucleus.

Transfer reactions are direct reactions which means that they proceed through a single step. The timescales involved for these reactions are of the order of $10^{-22} - 10^{-23}$ seconds which is roughly the time required for the projectile travelling at speeds close to the speed of light to travel the distance of dimensions of the radius of the target. This is in contrast to compound nuclear reactions where a compound nucleus is formed in the intermediate state and the reaction proceeds with no information about how that state was formed. The timescales for these reactions are also significantly higher as compared to transfer reactions. Only the outer valence nucleons take part in transfer reactions and rest of the core remains undisturbed. This makes them a useful tool to study single particle states.

Transfer reactions can be depicted as shown in Figure 1.2. They are denoted as either $[a + A \rightarrow b + B]$ or $[A(a, b)B]$. $A$ and $b$ are referred to as core nuclei as their constituent nucleon do not change their orientation throughout the reaction whereas $a$ and $B$ are called composite nuclei as they contain the exchanged nuclei in addition to the core. $A + a$ is referred to as the entrance channel of the reaction and $B + b$ as the exit channel. The kinematics of this reaction can be fully described by two-body kinematics since the initial as well as the final state contain only two bodies each.

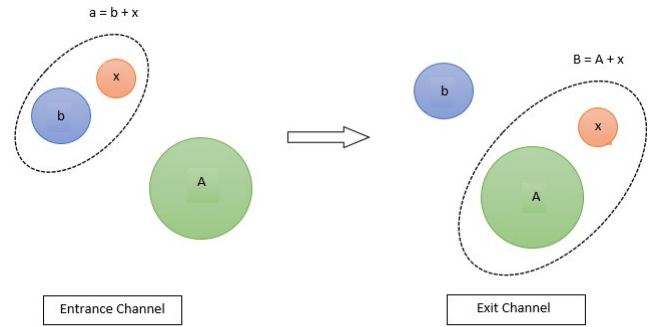

FIGURE 1.2: Schematic representation of Transfer Reactions

The most commonly used transfer reactions involve single nucleon transfer. This is particularly useful to avoid the complications arising from multi-nucleon transfer configurations. Moreover, the cross-sections for single nucleon transfer are much higher which is important to consider when performing reactions with radioactive ion beams. The transfer of a single neutron is more probable than that of a proton due to couloumbic forces. Reactions like $(d, p)$ and $(d, t)$ were carried out for this work which involve both the stripping and pick-up of a single neutron.

### 1.2.1 Inverse Kinematics

The $(d, p)$ and $(d, t)$ reactions have been used since a very long time to study the single particle structure of stable and long-lived isotopes using deuteron beams. The reactions carried out in this manner are said to be in normal kinematics since the target is much heavier than the beam. However, in recent times, when more exotic isotopes have to be studied, it is impossible to fabricate their targets since they are extremely short-lived. We then have to resort to inverse kinematics where deuteron is used as a target and the beam consists of the isotope to be studied. There arise several problems, of which two prominent ones are discussed below.

1. Inverse kinematics experiments have a large centre of mass motion. This leads to forward drift in laboratory frame and hence, less favourable angular coverages for ejectiles.

2. The beam currents for rare isotope beams used in inverse kinematics are significantly lower than that for stable beams. This enforces thick targets to be used to get measurable cross-sections for the reaction. This causes problems like huge energy losses for reaction products and poor energy resolution as the reaction can take place anywhere within the target.

Even after these problems, experiments in inverse kinematics experiments have made possible the spectroscopy of exotic isotopes and many useful results have been successfully achieved.

## 1.3 Angular Distributions Overview

The aim of angular distribution measurements is to estimate the multipolarity of $\gamma$-radiation from which, spin-parity assignments to excited nuclear states can be made (Ferguson 1965). It is simply a measurement of variation of $\gamma$-ray intensity measured as a function of angle between the direction of $\gamma$-ray emission and another fixed direction in space. This direction can be chosen in an experiment to be the direction of incident beam which populates the initial state or the direction of emission of another particle of even a $\gamma$-ray which aligns the decaying state.

Angular distribution measurements can be done in a laboratory by using a $\gamma$ detector of finite solid angle to count the no. of photons emitted in a given amount of time. These measurements can be repeated at different orientations of this detector with respect to the beam direction for similar times and intensity can be plotted as a function of $\theta$. This function is represented as

$$W(\theta) = \sum_{k=0}^{n} B_k P_k cos(\theta)$$

for reasons explained in Appendix A. These coefficients then depend on the total intensity of the transition as well as the angular dependence. To get rid of the total intensity dependence, these coefficients are normalized by dividing throughout by $B_0$. Thus the final expression reads as

$$W(\theta) = 1 + \sum_{k=1}^{n} a_k P_k cos(\theta) \tag{1.3}$$

where $a_k = \frac{B_k}{B_0}$. The final goal of these measurements is then to find the coefficients $a_k$'s which truly quantify the angular dependence.

## 1.4 Outline of Thesis

In summary, the goal of this work is to develop the technique for measuring angular distributions with our set-up. This will help to assign and confirm the spins of several states in current experiments as well as those to be carried out in future as we constantly push the limits for creating newer exotic beams for nuclear experiments. In this thesis we start, in Chapter 2, with the description of experiments carried out for this work [$^{95}$Sr(d,p)$^{96}$Sr; $^{95}$Sr(d,t)$^{94}$Sr]. The whole set-up including beam, target, detectors, etc. is described in detail. In the next chapter, we move to the theoretical model for calculating the expected angular distributions and discuss the effect of alignment of nuclei on these measurements. In Chapter 4, the techniques for the analysis of data and the results obtained are discussed. Methods used like Doppler Corrections, addback, headlight effect, etc. are . Finally, we conclude in Chapter 5 with the summary of this work and prospects for further applications.

# Chapter 2

# Experiments

## 2.1 General Overview

A detailed spectroscopy of States in $^{94}$Sr and $^{96}$Sr was carried out using the data from single nucleon transfer reactions in inverse kinematics. A 5.5 MeV/u $^{95}$Sr beam was impinged on a CD$_2$ target to look for single neutron stripping (d,t) and single neutron pickup (d,p) reactions, respectively. Both the reactions were carried out simultaneously using $\gamma$-particle coincidence techniques which was possible through a combination of position-sensitive particle detection (SHARC) and $\gamma$ detection (TI-GRESS). A schematic diagram of the experiment is shown in Figure 2.1.

These experiments were carried out at TRIUMF, Canada's national laboratory for nuclear and particle physics research during June 2014 with the Isotope Separator and Acceleration (ISAC-II) facility. This was one of the very first experiments exploring high mass region (A>30) with re-accelerated secondary beams and it marks an important step in the laboratory's ability to produce heavy ion exotic beams for performing nuclear reactions.

The various components in the experimental setup are described in brief throughout this chapter. Starting with beam delivery, the two main techniques for producing exotic beams are described followed by the description of beam delivery for this experiment. Subsequently, the deuteron target for this experiment is outlined. Finally, the detector systems for particle and $\gamma$ measurements are described.

## 2.2 Beam Delivery

The two main ways of producing radioactive ion beams (RIBs) or rare isotope beams, as they are called, are: In-flight fragmentation and Isotope Separation OnLine (ISOL). Both are complementary to one another and each technique has certain pros and cons. However, they often access different areas of nuclear chart and both types of facilities are required for exhaustively producing new exotic isotopes. Facilities like the NSCL in the US, RIKEN in Japan and GSI in Germany are in-flight fragmentation

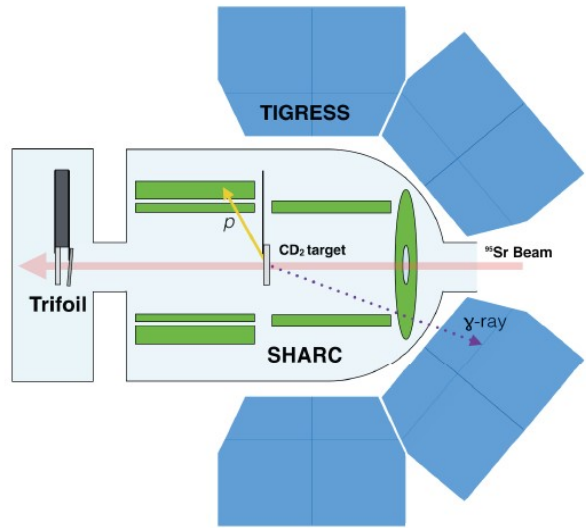

FIGURE 2.1: Schematic diagram of the experiment showing the detector set-up

facilities while TRIUMF in Canada and ISOLDE at CERN are ISOL facilities.

The in-flight fragmentation facility uses a high energy (>50 MeV/u) heavy-ion beam impinged upon a thin target which results in very less energy loss in the production target. The heavy ion undergoes fragmentation as a result of interaction with the target and produces a fast radioactive ion cocktail beam. The main advantage of this method is that we can study extremely short-lived isotopes with half-lives up to an order of a nanosecond as the initial heavy-ion beam energy is carried forward by the RIB. However, this requires highly efficient mass separation and generally, the energy distribution of the beam is quite broad. The cross-sections also tend to be low which results in poor beam quality. In contrast to this, the ISOL facility used a light ion beam, usually proton or $\alpha$ particles impinged upon a thick target. The primary beam looses all of its energy in the production target which maximizes the production yield. However, it is challenging to efficiently extract the exotic ions produced in the target. It is for this reason that the production targets are operated at temperatures of several thousand Kelvin to diffuse the exotic ions to the surface. The secondary beams have to be re-accelerated to perform nuclear experiments which result in additional inefficiencies. Also, the isotopes with half-lives less than 10 ms cannot be produced currently with this approach. However, we can get good quality beams with precise final energies.

For these experiments, the TRIUMF 500 MeV cyclotron was used to produce a high intensity beam of protons with beam current of upto $10\mu$A. The beam was then sent to ISAC facility where it impinged on a thick Uranium Carbide ($UC_x$) target. Proton-induced uranium fission and spallation within the target produced a yield consisting of a wide variety of nuclei. These isotopes were extracted from the $UC_x$ target and numerous stripping and filtering techniques were applied to optimize the purity and rate of $^{95}$Sr beam.

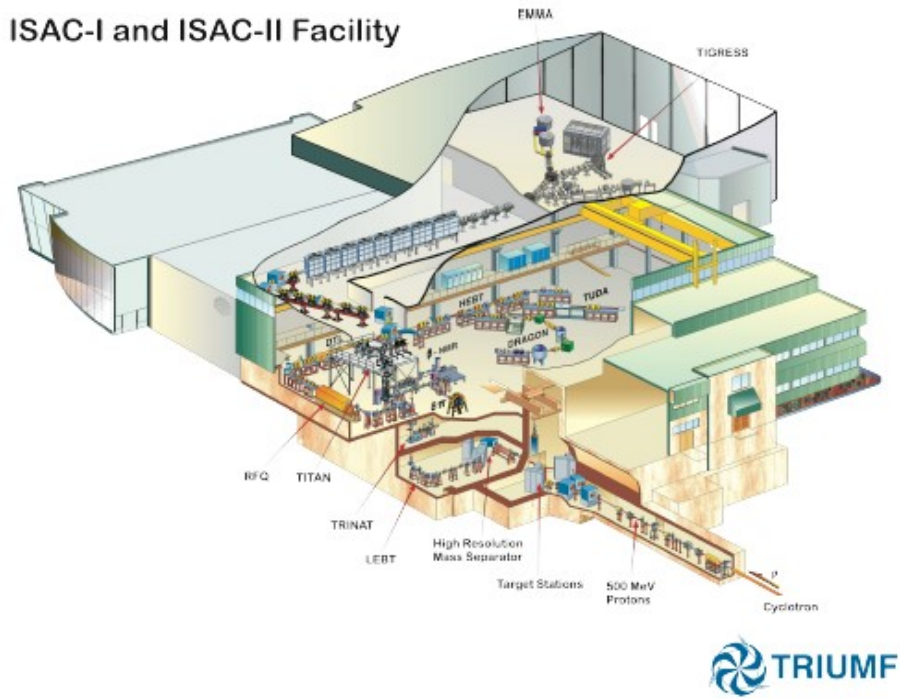

FIGURE 2.2: Diagram of TRIUMF ISAC Facility. The $^{95}$Sr$^{16+}$ beam was delivered to TIGRESS Experimental Station in ISAC-II.

Surface ions were ionized into singly charged (1$^+$) state using TRIUMF Resonant Ionization Laser Ion Source (TRILIS). TRILIS uses a high intensity laser beam to ionize atoms using multistep resonant photon absorption which leads to efficient and element selective ionization. TRILIS was used to enhance the extraction rate of $^{95}$Sr isotopes as compared to other contaminants which are also produced in the production target. The cocktail beam was then sent to ISAC Mass Separator and later fed to the Charge State Booster (CSB). Here, the isotopes were charge-bred by an Electron Cyclotron Resonance (ECR) plasma to 16$^+$ charge state. This enabled the $^{95}$Sr isotopes to be selectively accelerated in the Radio-Frequency Quadrupole (RFQ) by tuning the time-dependent electric fields. Following the RFQ, the contaminants were deflected out of the beam using the bending dipole magnets in the accelerator chain. This highly charged $^{95}$Sr beam was re-accelerated to 524 MeV (5.515 MeV/u) using the superconducting linear accelerator before being finally delivered to the TIGRESS Experimental Station in ISAC-II. An overview of ISAC-I and ISAC-II facilities is shown in Figure 2.2.

## 2.3 Target

The deuterated polyethylene was used as a deuteron target in these experiments. The target, however, wasn't fully deuterated. Elastic scattering measurements with deuteron and hydrogen (d,d) and (p,p) were used to determine the target deuteration which was found to be 92(1)%. The nominal thickness was chosen to be 5.0 $\mu$m during manufacturing as a trade-off between total reaction cross-section and energy broadening effects. Thicker targets produce high yields but the reaction can take place anywhere in the target. So, there is substantial energy loss for some nuclei and none for others giving broad energy distributions. The target thickness measurements were performed by using a triple-$\alpha$ source containing $^{239}$Pu, $^{241}$Am, $^{244}$Cm. Knowing the deuteration factor, energy loss of $\alpha$ particles were measured and thickness was determined as the difference between their range in the target material before and after the target. The experimentally measured value was 4.4(4) $\mu$m.

## 2.4 Detector Systems

The following section describes the various detectors used for the experiments described in this thesis along with their operational principles.

### 2.4.1 Auxiliary Detectors

In addition to the detector systems for reaction products, two other auxiliary detectors are used to monitor the beam. TBragg is located upstream before the beam interacts with the target whereas Trifoil is a counter located downstream from the target. Both of these are discussed in brief in this section.

**TBragg**

The TBragg is a gas-filled ionization chamber used for identifying heavy-ions present in the radioactive cocktail beam. As the beam enters the gas chamber, it ionizes the gas producing free ions and electrons. The pressure of the gas is chosen so that the beam is completely stopped in the chamber and full energy measurements can be made. The electrons are drifted towards the anode owing to the longitudinal electric field. The strength of this field determines how quickly and efficiently electrons can be collected. A pulse-shaped analysis of the electrical signal is used to extract information about the incident ion: the slope of the signal is related to the stopping power of the ion whereas the total amplitude is related to the incident energy of the ion. Figure 2.3 shows the components of mass 95 beam in the TBragg spectrometer.

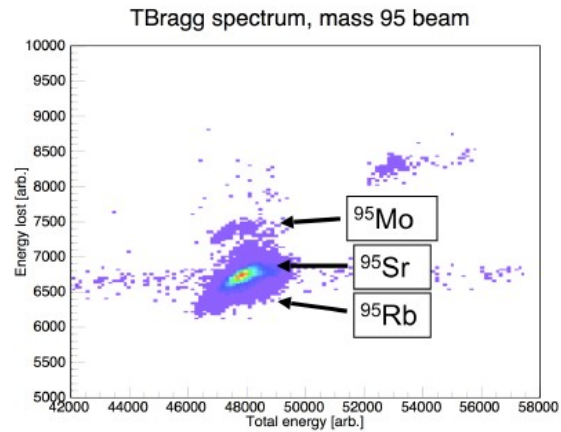

FIGURE 2.3: Beam identification plot of mass 95 beam

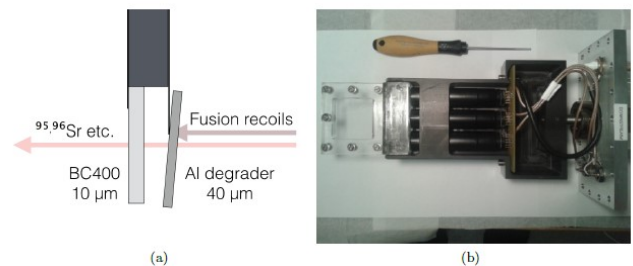

FIGURE 2.4: (a) Schematic diagram of trifoil (b) Photograph of trifoil during set-up

It indicates $98.5(5)\%$ purity of the beam with other isobaric contaminants.

**Trifoil**

The Trifoil scintillator is an auxiliary system consisting of BC400 foils connected to a set of three photomultipliers. Due to the very fast scintillating light signals from the BC400 foil, it can be used as a RIB counter. It is set-up downstream of the target chamber as shown in Figure 2.4. During the reaction, the mass 95 radioactive beam can undergo fusion with the target and quickly evaporate several light particles into SHARC. This can interfere with our reaction products and make the analysis more difficult. This problem can be solved by using an Aluminium degrader in front of trifoil which blocks the heavy fusion products. Thus only the reaction products can reach trifoil and the signal can be used in coincidence with those from our main detector systems.

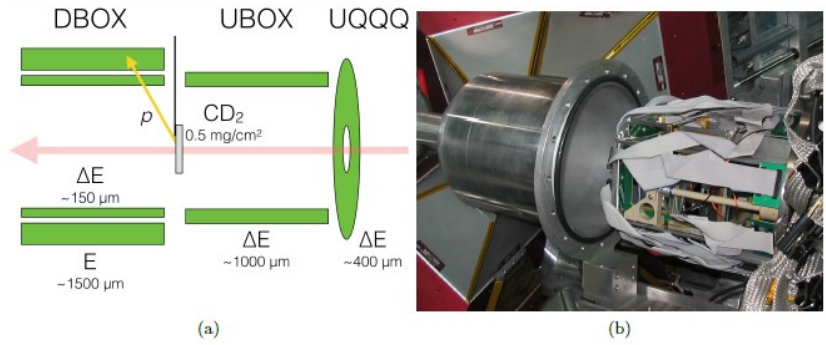

FIGURE 2.5: (a) Schematic Diagram of SHARC (b) A photograph of SHARC detector being installed

### 2.4.2 SHARC

SHARC stands for Silicon Highly-segmented Array for Reactions and Coulex (Coulomb Excitation). It is an array of Double-Sided Striped Silicon Detectors (DSSD) with acceptance of almost full $4\pi$ in its full instrumentation. It has excellent angular resolution which is very useful for making angular distribution measurements. Figure 2.5 shows a schematic diagram of SHARC as it was used in the experiment.

SHARC detectors were primarily used to detect protons and tritons from the (d,p) and (d,t) reactions. The $\gamma$-rays detected in coincidence with protons were assigned as coming from the excited states of $^{96}$Sr while those in coincidence with tritons were from the excited states of $^{94}$Sr. One of the operational principle for SHARC detectors used for detecting tritons is the $\Delta$E-E counter Telescope System. It involves the measurement of energy loss in a thin counter. From the Bethe-Bloch equation for stopping power (Knoll 2010),

$$-\frac{dE}{dx} = Kz^2 \frac{Z}{A} \varrho \frac{1}{\beta^2} \left[ \frac{1}{2} \ln \frac{2m_e c^2 \beta^2 \gamma^2 T_{max}}{I^2} - \beta^2 - \frac{\delta}{2} \right] \tag{2.1}$$

we see that the product $\Delta$E and E is equal to $kz^2M$ (assuming that the factors in the bracket will be constant for non-relativistic case), where $ze$ is the charge of the particle, $M$ is its mass and $k$ is a constant depending on the absorber material. Plotting the value of energy loss in thin counter, $\Delta E$ against the total energy $E$ gives us a family of mass hyperbolas corresponding to different values of $z^2m$ which in turn gives us the particle identification (PID). Figure 2.6 shows one such measurement. Another method is the time-of-flight method which is similar to this method but uses TAC to measure the time taken by particles to fly between two counters and then plot the mass hyperbolas.

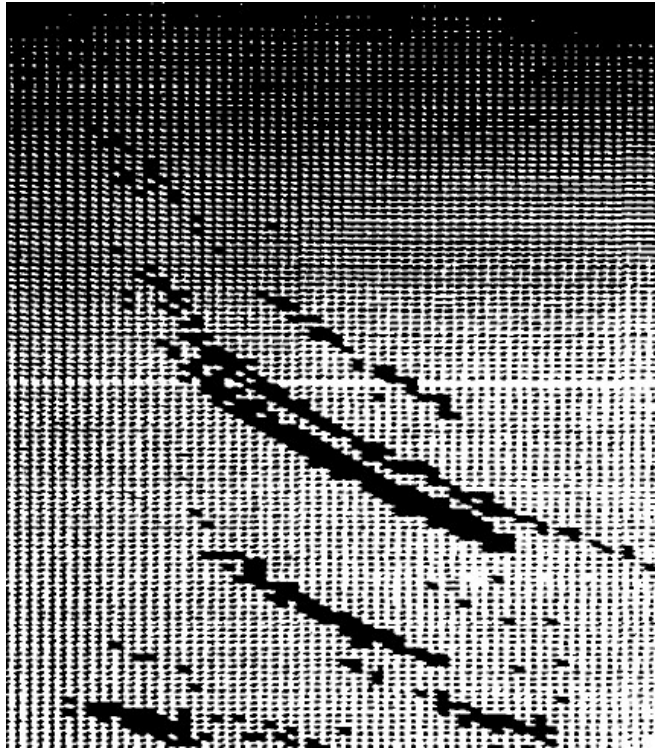

FIGURE 2.6: $\Delta E - E$ measurement showing mass hyperbolas for $^{16}$O, $^{13}$C ,$^{12}$C,$^{7}$Li, $^{6}$Li, $^{4}$He (Krane and Halliday 1988).

### 2.4.3 TIGRESS

TIGRESS stands for TRIUMF-ISAC Gamma-Ray Escape Suppressed Spectrometer which is an array of High-purity Germanium (HPGe) Clover detectors. The main advantage of these detectors is the excellent energy resolution which is of prime importance in $\gamma$-ray spectroscopy as the nuclei can have tightly spaced energy levels. TIGRESS is purpose-built for use in reaction studies where photons are emitted from recoiling nuclei. Hence, excellent angular resolution is a must feature in these types of detectors for precise Doppler Correction.

Each TIGRESS Clover detector is made up of four closed-ended n-type coaxial HPGe crystals. The individual crystals have an eight-fold segmentation: four quadrants and a lateral divide which produces a 32-fold segmentation in a single clover. This gives very good angular resolution so that precise Doppler Correction can be made. Good angular resolution also helps to improve the quality of our data through addback algorithm which is described in the next chapter. Figure 2.7 shows the segmentation within a single clover.

The clovers are arranged into constant $\theta$ rings with 4 clovers each at $45 \deg$ and $135 \deg$ and 8 clovers at $90 \deg$ with respect to the beam axis. There is close to full $\phi$ coverage in each of these rings. In this experiment, $45 \deg$ ring was excluded since the additional space was required for SHARC pre-amplifiers and the remaining 12 clovers were used. Each clover is surrounded by four Bismuth Germanate (BGO) Compton Suppressor Shields which can be used as a veto for Compton-Scattered γ-rays. That is whenever there is a hit in the BGO shields, we discard that event since full energy is not deposited in the crystal. This greatly reduces the background and makes the data clean. Figure 2.8 shows a photograph of TIGRESS array as it was used in this work.

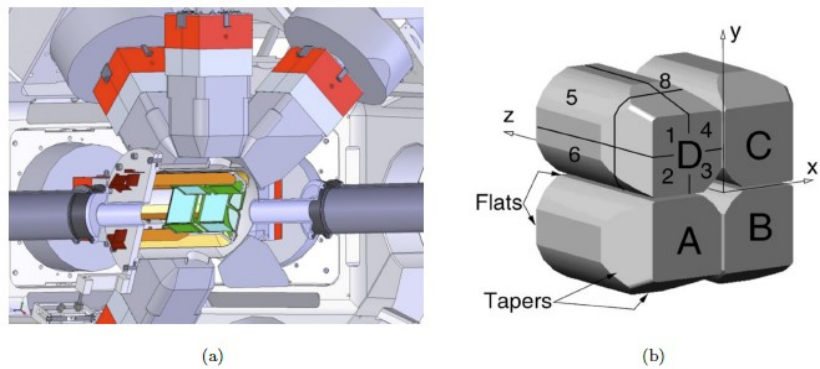

FIGURE 2.7: (a) CAD cutaway drawing of TIGRESS (b) Segmentation of TIGRESS Clover

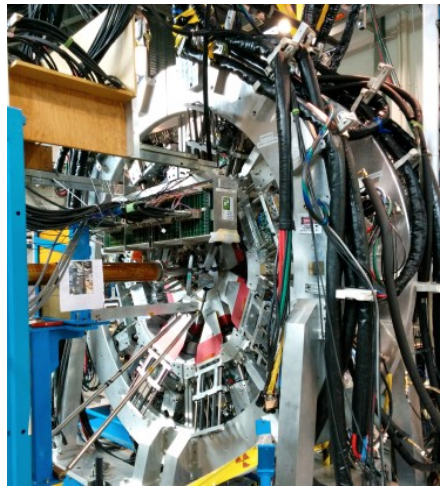

FIGURE 2.8: Installation of TIGRESS in Experimental Setup

There are two operational modes for the array: optimized peak-to-total and high-efficiency. In the optimized peak-to-total mode, the BGO shields are brought forward so that they are flush with γ-rays at the front face of the detector. This induces maximum reduction in Compton-scattered events. In the high-efficiency mode, the BGO shields are pulled back and the crystals form a continuous face. This greatly increases the solid angle coverage and hence the overall efficiency of the detector.

Each TIGRESS clover contains an in-built pre-amplifier and high voltage supply for each segment and also one for the core. These are all connected to a shared cryostat which is maintained at 77K using a liquid Nitrogen LN$_2$ reservoir.

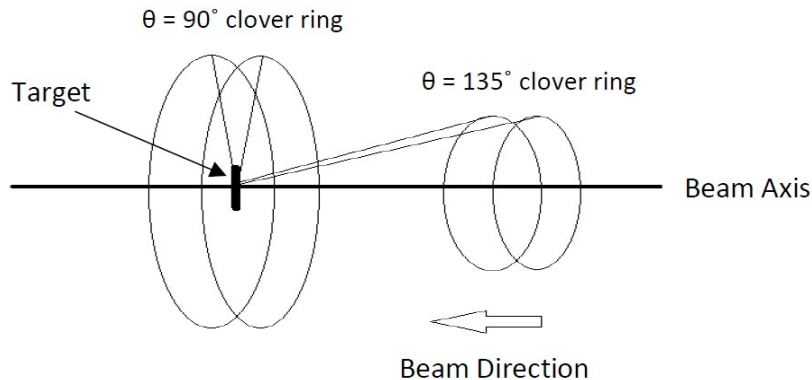

FIGURE 2.9: Four arrays of constant angle rings for Angular Distribution Measurements

## 2.5 Array of Constant Angle Rings for Angular Distribution Measurements

Angular Distribution measurements demand position sensitive detection of $\gamma$-rays. Since TIGRESS contains decent solid angle coverage with excellent angular resolution, it makes a very good system to perform these measurements. Each segment can be thought of as a stand-alone detector at a different angle and measuring the intensities of particular transitions can give us the angular distributions. However, the statistics at the segment level are very poor to identify the peak and measure its intensity. As a result, we perform our analysis at the crystal level with much better statistics.

However, all the crystals do not correspond to different angles. Each clover has four crystals which form groups of two, corresponding to a constant angle with respect to the beam axis. Thus, all the crystals from a particular clover ring are grouped into two constant angle rings of detectors on each side of the clover ring. The two clover rings used in this experiment give us a total of four constant angle rings of crystals as shown in Figure 2.9. We can thus perform angular distribution measurements with four data points and much better statistics as compared to the crystal level.

# Chapter 3

# Theory

This chapter includes the theoretical formulation of $\gamma$-ray angular distributions. A brief overview of $\gamma$-ray transitions and the selection rules is given followed by the conditions for anisotropic angular distributions. Thereafter, the alignment conditions and its effect on the $a_2$ and $a_4$ coefficients is described. Finally, the theory for mixing of multipolarities is described with a technique for finding the experimental mixing ratio.

## 3.1 $\gamma$-Transitions

$\gamma$-radiation is one of the most abundant form of radioactive decay. It often follows other decay modes which leave the daughter nucleus in it's excited state. The energy difference between the two level is emitted as a $\gamma$-ray photon of the corresponding energy. These photons have a total angular momentum associated with them which defines the multipolarity of the transition. Each nuclear state is also characterized by it's total angular momentum and law of conservation of angular momentum leads us to the first selection rule for $\gamma$-ray emission whereas the law of conservation of parity gives the second selection rule. Let the total angular momentum and parity of the initial and final states be represented by $J_i, \pi_i$ and $J_f, \pi_f$ respectively, and let $l$ be the multipolarity of the emitted $\gamma$-ray. Then the parity of the operator for electric transition is $(-1)^l$ and for magnetic transitions is $(-1)^{l+1}$. The selection rules can be represented as (Krane and Halliday 1988):

1. $|J_i - J_f| \leq l \leq (J_i + J_f)$, so that they satisfy the triangle inequality and the total angular momentum difference between the two levels is carried away by the $\gamma$ photon.

2. $\pi_i \pi_f = -1$ for odd electric and even magnetic multipoles.
   $\pi_i \pi_f = 1$ for even electric and odd magnetic multipoles.

When either of $J_i$ or $J_f$ is $0$, the $\gamma$-transition has a unique multipolarity and there cannot be any mixing whatsoever. However, when both $J_i$ and $J_f$ are $0$, the selection rules demand an $l = 0$ transition which is not possible since the photon is an elementary particle with inherent spin of $1$ and cannot have a total angular momentum less than $1$. Hence, such transitions proceed through internal conversion electrons which are often accompanied by the emission of X-rays as the electrons cascade down to the empty place left by the conversion electron. A stretched transition is one in which the photon carries angular momentum equal to the algebraic difference between the angular momentum of initial and final state i.e. all three vectors are collinear. Usually, the lowest permitted multipoles are the most preferred modes of emission since the expected intensity falls down by a factor of approximately $10^{-5}$ with each increasing order of multipole.

## 3.2 Angular Distributions Theory

Angular distributions are quantified by the coefficients $a_k$'s as discussed in Section 1.3. They are attributed to preferential population of magnetic sub-states of nuclear levels. If all the sub-states are equally occupied, the transition will be isotropic. This can be understood from the completeness condition of Poynting vector functions ($F_J^m(\theta)$). They represent the energy-flow in a transition as a function of angle ($\theta$) (Hagström, Nordling, and Siegbahn 1965) and follow the property

$$\sum_{-m}^{m} F_J^m(\theta) = 1$$

Let us take for example the dipole transition $1^+ \rightarrow 0^+$. If all the magnetic sub-states in $J_i = 1$ state are equally occupied, then the distribution is isotropic. However, if the sub-state $m = 0$ is preferentially populated, then $F_1^0 = 3sin^2(\theta)$ which gives us a $sin^2(\theta)$ dependence for this dipole transition. Similarly, it gives a $cos^2(\theta)sin^2(\theta)$ dependence for a quadrupole transition ($2^+ \rightarrow 0^+$) when only the $m = 0$ sub-state is populated.

Let us define a population parameter $w(m_i)$ such $\sum_i w(m_i) = 1$. Then the case for equal sub-state population will be defined by $w(m_i) = \frac{1}{2J+1}$ for all $i$. There are departures form this case that gives anisotropic angular distributions, viz. polarization and alignment. Polarization is when ($w(m_i) \neq w(-m_i)$) whereas alignment is when $w(m_i = w(-m_i) \neq \frac{1}{2J+1}$ for some values of $i$. The following three conditions are then quite apparent:

1. Transitions from spin 0 state will always be isotropic since it can neither be aligned nor polarized.

2. Spin $\frac{1}{2}$ state cannot be aligned and transitions from those will only be anisotropic if the state is polarized.

3. States with spin greater than 1 can either be aligned or polarized.

The next section defines a more rigorous alignment parameter, the statistical alignment tensor and gives a theoretical formalism for obtaining the expected values of the coefficients $a_k$'s given the alignment, spins of initial and final states and the multipolarity of transition.

### 3.2.1 Theoretical Formalism

The angular distribution coefficients for a pure transition can be represented in terms of a statistical alignment tensor, which as the name suggests depends upon the alignment of decaying state and a coupling parameter which depends on the angular momenta of participating states and the multipolarity of $\gamma$-transition. The statistical tensor is defined as

$$\rho_k(J) = \sqrt{2J+1} \sum_m (-1)^{J-m} < J, m; J, -m | k, 0 > P_m(J) \qquad (3.1)$$

where $P_m(J)$ can written as

$$P_m(J) = \frac{\exp(-\frac{m^2}{2\sigma^2})}{\sum_{m'=-J}^{J} \exp(-\frac{m'^2}{2\sigma^2})} \qquad (3.2)$$

Thus, we see that $\sigma$ is the only free parameter describing the alignment of decaying state which is approximated to be a Gaussian distribution about $m = 0$ state. In practice, this is quite accurate representation since $m = 0$ state is most likely to be populated. The coupling constant $F_k(J_f L L J_i)$ can be written as

$$F_k(J_f L L J_i) = (-1)^{J_f - J_i - 1} [(2L+1)^2 (2J_i+1)]^{\frac{1}{2}} < L, 1; L, -1 | k, 0 > W(J_i J_i L L; k J_f) \qquad (3.3)$$

where $< \ldots | \ldots >$ is the Clebsch-Gordon Coefficient and $W(\ldots; \ldots)$ is the Racah-Coefficient. The values for $F_k$'s for different combinations of $J_i$, $J_f$ and $L$ were read from the Angular Distribution tables by Yamazaki (1966). The angular distributions for unmixed transitions can then be represented as

$$A_k(J_i L L J_f) = \rho_k(Ji) F_k(J_f L L J_i) \qquad (3.4)$$

When two multipolarities $L_1$ and $L_2$ mix together by an amount $\delta$ such that

$$\delta = \frac{< J_f || L_2 || J_i >}{< J_f || L_1 || J_i >}$$

the angular distribution coefficients can be written as

$$A_k(J_i L_1 L_2 J_f) = \rho_k(Ji) \frac{1}{1+\delta^2} \{ F_k(J_f L_1 L_1 J_i) + 2\delta F_k(J_f L_1 L_2 J_i) + \delta^2 F_k(J_f L_2 L_2 J_i) \} \qquad (3.5)$$

The statistical parameter is still as defined earlier, and the coupling constants become

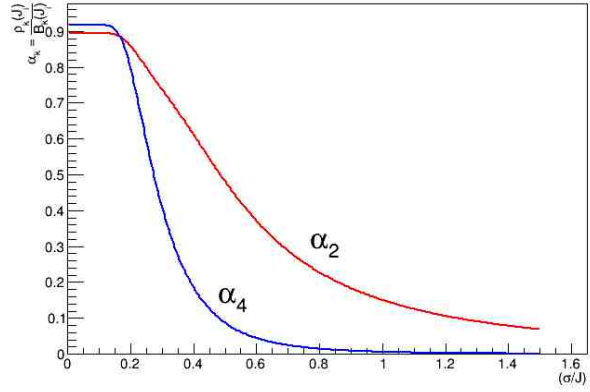

FIGURE 3.1: Attenuation factors at varying alignment parameters for $J = 2$

$$F_k(J_f L_1 L_2 J_i) = (-1)^{J_f - J_i - 1}[(2L_1+1)(2L_2+1)(2J_i+1)]^{\frac{1}{2}} < L_1, 1; L_2, -1|k, 0 > W(J_i J_i L_1 L_2; kJ_f) \tag{3.6}$$

The definition of our statistical tensor is invalid for full alignment i.e. when only the $m = 0$ state is populated since $\sigma = 0$ in this case. Hence, for full alignment, it can be defined as

$$B_k(J_i) = \sqrt{(2J+1)}(-1)^J < J, 0; J, 0|k, 0 > \qquad \forall \text{ integral spin}$$
$$B_k(J_i) = \sqrt{(2J+1)}(-1)^{J-\frac{1}{2}} < J, \frac{1}{2}; J, -\frac{1}{2}|k, 0 > \qquad \forall \text{ half-integral spin} \tag{3.7}$$

Then, for this ideal case,

$$A_k^{max}(J_i L_1 L_2 J_f) = \frac{1}{1+\delta^2}\{f_k(J_f L_1 L_1 J_i) + 2\delta f_k(J_f L_1 L_2 J_i) + \delta^2 f_k(J_f L_2 L_2 J_i)\} \tag{3.8}$$

where

$$f_k(J_f L_1 L_2 J_i = B_k(J_i)F_k(J_f L_1 L_2 J_i)$$

In actual cases when the alignment is partial,

$$A_k(J_i L_1 L_2 J_f) = \alpha_k(J_i)A_k^{max}(J_i L_1 L_2 J_f) \tag{3.9}$$

where

$$\alpha_k(J_i) = \frac{\rho_k(J_i)}{B_k(J_i)}$$

The variation of coefficients $a_2$ and $a_4$ with increasing alignment for $J = 2$ state is shown in Figure 3.1. In reality, the coefficients $a_k$'s are further attenuated because of the fact that the detectors we use for counting the $\gamma$-rays have finite solid angle coverage. Thus, all the points on the detector face do not correspond to similar $\theta$. An easy way to combat this effect, as per Krane (1972), is that the whole detector area is divided into infinitesimal patches and the coefficients are calculated for each patch with the corresponding angle $\theta$. Finally, the results are integrated and compared to the coefficients assuming constant $\theta$. Calculations are done using FORTRAN code as described in Krane (1972). The final result is additional coefficient attenuation factors $\beta_k$'s such that the above equation becomes

$$A_k(J_i L_1 L_2 J_f) = \alpha_k(J_i)\beta_k A_k^{max}(J_i L_1 L_2 J_f) \tag{3.10}$$

The inputs to run the computer code include geometrical parameters of the detector system and the $\gamma$-ray absorption coefficient in the detector material. Plugging the values for our system, we find that

$$\beta_2 = 0.9718$$

$$\beta_4 = 0.9080$$

It should be noted that higher order coefficients attenuate faster than lower order coefficients.

In the case of angular correlation measurements when the $\gamma$ is emitted following another $\gamma$ in a cascade, the statistical alignment parameter of the initial state for both the transitions are related. If the state $J_f$ is formed only through the preceding transition $J_i \rightarrow J_f$, the statistical tensor for state $J_f$ is expressed in terms of that of state $J_i$ as follows:

$$\rho_k(J_f) = U_k(J_i L_1 L_2 J_f)\rho_k(J_i) \tag{3.11}$$

where,

$$U_k(J_i L_1 L_2 J_f) = \frac{1}{1+\delta^2}[u_k(J_i L_1 J_f) + \delta^2 u_k(J_i L_2 J_f)] \tag{3.12}$$

$$u_k(J_i L_1 J_f) = (-1)^{J_i + J_f - L_1}\sqrt{(2J_i+1)(2J_f+1)}W(J_i J_i J_f J_f; kL_1) \tag{3.13}$$

Therefore,

$$\alpha_k(J_f) = U_k(J_i L_1 L_2 J_f)\frac{B_k(J_i)}{B_k(J_f)}\alpha_k(J_i) \tag{3.14}$$

This sums up the theoretical formulation used for calculating the expected angular distribution coefficients in this work.

### 3.2.2 Experimental Mixing Ratio

As we can see from Equation 3.5, the angular distribution coefficients depend on alignment $\sigma$ and the mixing ratio $\delta$. $\sigma$ can be measured experimentally from the experimental angular distribution coefficients for unmixed transitions and using Equation 3.4. Once we know the value of $\sigma$ for a state which is constant for a given reaction channel, we can plot a graph of $\chi^2$ vs $\delta$ where $\chi^2$ is given by

$$\chi^2 = \frac{[a_2(expt) - a_2(theo)]^2}{3[\Delta a_2(expt)]^2} + \frac{[a_4(expt) - a_4(theo)]^2}{3[\Delta a_4(expt)]^2}$$

Where $\Delta a_{2,4}$ are the uncertainties in the experimentally observed angular distribution coefficients. $a_{2,4}(theo)$ are calculated using Equation 3.5 keeping $\delta$ as a free parameter so that $\chi^2$ is a function of $\delta$. Minimizing this $\chi^2$ gives us the value of experimental mixing ratio for that transition. This is known as the $\chi^2$ minimization procedure which was given by Singh (1992). The error in this quantity can be found by taking the value of $\delta$ at $\chi^2_{min} \pm 1$. In practice, the graph is plotted against $tan^{-1}\delta$ as opposed to $\delta$ so that the nature of the graph is cyclic and it is easy to trace the minimum value.

## 3.3 Alignment in Transfer Reactions

The alignment in reactions involving beam impinging on a target can be approximated to be Gaussian in nature. This is because there is a large probability for the $m = 0$ substate to be populated followed be decreasing probability for occupancy of higher $m$-substates. The exact values for the occupation probability can, however, be calculated by a rather simple analysis involving Clebsch-Gordon coefficients and $\sigma$ can be estimated from those values. More information on the Clebsch-Gordon coefficients and their physical interpretation is given in Appendix B. They represent the probability amplitude of finding a particular configuration of spins in a total spin state. The square of these coefficients, hence, represent the occupation probability. In this section, this type of analysis is presented for the directly populated $2_1^+$ state in $^{96}$Sr and results for similar analysis of other states is presented.

Consider the reaction $^{95}$Sr(d,p)$^{96}$Sr with reference to Figure 3.2. Among the reactants, deuteron has spin 1 and $^{95}$Sr has ground state spin-parity of $\frac{1}{2}^+$. We are interested in finding the relative occupation probability of different $m$ substates. Deuteron with Spin 1 initially can have $m = 0, \pm 1$ with equal probabilities since the target is unpolarized. Upon impinging $^{95}$Sr on it the total spin of the system will be $\frac{3}{2}$ or $\frac{1}{2}$ with the possible m substates of $m = \pm\frac{3}{2}, \pm\frac{1}{2}$. The relative occupation of these states can be found as follows:

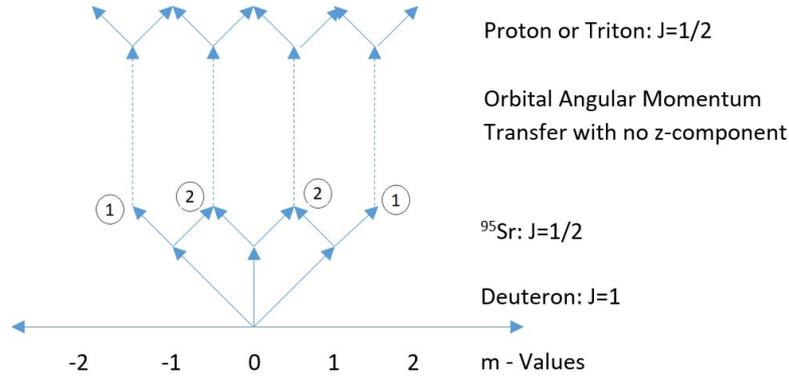

FIGURE 3.2: Alignment of $J = 2_1^+$ state in $^{96}$Sr directly populated
through $^{95}$Sr(d,p)$^{96}$Sr reaction.

- For $m = \frac{3}{2}$:

  Occupation probability = $< 1, 1; \frac{1}{2}, \frac{1}{2} | \frac{3}{2}, \frac{3}{2} >^2 = 1$

- For $m = \frac{1}{2}$:

  Occupation probability = $< 1, 0; \frac{1}{2}, \frac{1}{2} | \frac{3}{2}, \frac{1}{2} >^2 + < 1, 0; \frac{1}{2}, \frac{1}{2} | \frac{1}{2}, \frac{1}{2} >^2 + < 1, 1; \frac{1}{2}, -\frac{1}{2} | \frac{3}{2}, \frac{1}{2} >^2$
  $+ < 1, 1; \frac{1}{2}, -\frac{1}{2} | \frac{1}{2}, \frac{1}{2} >^2 = 2$

The corresponding occupation probabilities for the negative $m$-values will be the same owing to the symmetry of the problem. Thus, at this stage, the $m = \pm\frac{1}{2}$ states are twice as likely to be populated as $m = \pm\frac{3}{2}$ state.

Next, we take into account the orbital angular momentum imparted to the system by the beam. Theoretically, the maximum angular momentum it can impart to the system is calculated to be $14\hbar$ units by semiclassical calculations using $\vec{L} = \vec{r}$ x $\vec{p}$. The radius of $^{95}$Sr can be calculated by $R = R_o A^{\frac{1}{3}}$ and the momentum can be calculated from the beam energy. However, we encounter the states with maximum total angular momentum of $4\hbar$ in our analysis. Hence, we can safely assume that states with all spin values are present and only consider those which lead us to the required spin value for our state under consideration. The orbital angular momentum imparted by the beam will lie in the $xy$ plane as depicted in Figure 3.3. Hence, there will be no z-component for this part and the occupation values for m-substates remain unchanged. At this stage in our analysis, we have $J = \frac{1}{2}, \frac{3}{2}, \frac{5}{2}, \ldots$ with the maximum z-component of $m = \frac{3}{2}$.

In the last step, a free proton (or triton) is emitted with a ground state spin of $\frac{1}{2}$. This has to be included in our m-diagram to calculate the final occupation probabilities of our required states. While doing so, the results of the previous stage have to be taken into consideration where $m = \pm\frac{1}{2}$ state is twice likely as populated as $m = \pm\frac{3}{2}$ state. For the $J = 2$ state, the analysis of possible m-substates is as follows:

- For $m = -2$:

  Occupation probability = $< \frac{5}{2}, -\frac{3}{2}; \frac{1}{2}, -\frac{1}{2} | 2, -2 >^2 + < \frac{3}{2}, -\frac{3}{2}; \frac{1}{2}, -\frac{1}{2} | 2, -2 >^2 = \frac{7}{6}$

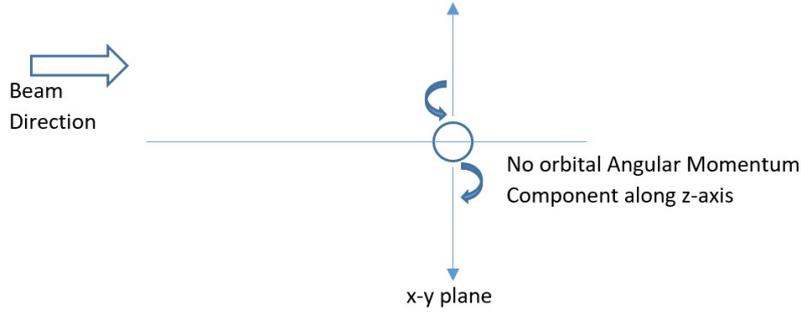

FIGURE 3.3: Orbital angular momentum transferred by the beam to the system having no $z$-component.

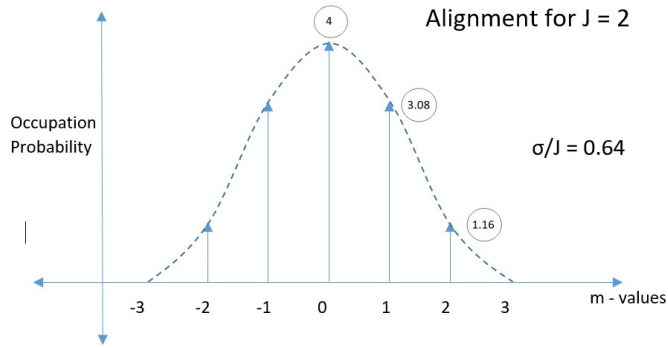

FIGURE 3.4: Occupation probabilities of different m=substates for $J = 2^+_1$ state.

- For $m = -1$:

  Occupation probability = $< \frac{5}{2}, -\frac{3}{2}; \frac{1}{2}, \frac{1}{2} | 2, -1 >^2 + < \frac{3}{2}, -\frac{3}{2}; \frac{1}{2}, \frac{1}{2} | 2, -1 >^2 + 2 * [< \frac{5}{2}, -\frac{1}{2}; \frac{1}{2}, -\frac{1}{2} | 2, -1 >^2 + < \frac{3}{2}, -\frac{1}{2}; \frac{1}{2}, -\frac{1}{2} | 2, -1 >^2] = \frac{37}{12}$

- For $m = 0$:

  Occupation probability = $2 * [< \frac{5}{2}, \frac{1}{2}; \frac{1}{2}, -\frac{1}{2} | 2, 0 >^2 + < \frac{3}{2}, \frac{1}{2}; \frac{1}{2}, -\frac{1}{2} | 2, 0 >^2 + < \frac{5}{2}, -\frac{1}{2}; \frac{1}{2}, \frac{1}{2} | 2, 0 >^2 + < \frac{3}{2}, -\frac{1}{2}; \frac{1}{2}, \frac{1}{2} | 2, 0 >^2] = 4$

These are plotted as in Figure 3.4. The value of $\sigma$ is calculated from here using the formula for standard deviation

$$\sigma = \sqrt{\frac{\sum_{i=1}^{n}(x_i - \bar{x})^2}{n}} \qquad , \bar{x} = \frac{\sum_{i=1}^{n} x_i}{n}$$

Following this analysis, the value of $\sigma/J$ for the directly populated $J = 2$ state in our experiments is calculated to be $\sigma/J = 0.6367$. Following a similar analysis for $J = 1$ and $J = 3$ state, we get the $\sigma/J$ values as $0.4330$ and $0.4358$ respectively. By this method, knowing about the $\gamma$-cascade sequence, we can even find the alignment for subsequent states through Equation 3.11. This way, we can precisely know the alignment of all levels and can accurately calculate the angular distribution coefficients.

# Chapter 4

# Analysis and Results

In this chapter, the analysis procedure will be outlined. The analysis was carried out using the GRSISort Software package (Bender, Bildstein, and Dunlop 2012) based on the ROOT Framework. Details of the specific programs used and developed for this work are outlined in Appendix C.

## 4.1 TIGRESS

### 4.1.1 Calibrations

A $^{152}$Eu $\gamma$-ray source was used to calibrate the energies and efficiencies of TIGRESS Array. The energies and intensities of the strongest transitions as per the data from National Nuclear Data Center (NNDC) were used for the same. The main advantage of using this source is the vast range of energies of $\gamma$ emissions from this source ranging from around 100 keV up to 1400 keV which is largely overlapping with our region of interest for this work. Figure 4.1 shows the corresponding energy spectra in a single crystal. The energy calibrations were done using a linear fit which was sufficiently accurate for our purpose.

The relative efficiencies at each energy were calculated by dividing the total number of counts in each peak by the relative intensities of each transition. In principle, these efficiencies would be different for each crystal or a constant angle array for the purpose of our analysis. Calibration spectra at crystal level were added to get those at constant angle array and relative efficiencies at each such array are calculated for each energy along with the corresponding errors.

### 4.1.2 Doppler Corrections

The reaction products will be recoiling due to the incident beam energy. $\gamma$-rays from these recoiling nuclei will be Doppler shifted due to the motion of their source. This effect depends on the energy of $\gamma$ photon emitted, the speed of recoiling nuclei and the angle between the motion of source and $\gamma$ photon emission ($\theta$). The measured

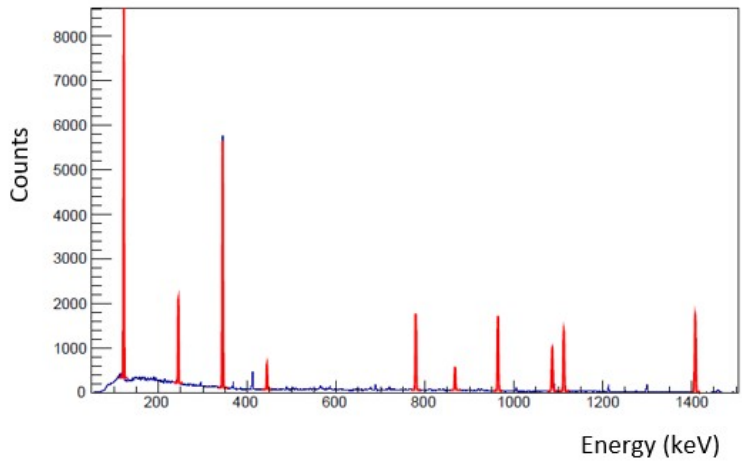

FIGURE 4.1: $^{152}$Eu $\gamma$-ray source spectrum for a single crystal.

energy $E'$ is then related to the energy of emission in rest frame of recoiling nuclei as

$$E = E'\gamma(1 - \beta cos(\theta))$$

Where $\beta$ and $\gamma$ are the usual Lorentz factors encountered in relativity. The velocity of recoiling nuclei was calculated on an event by event basis by reaction kinematics and hence, the Lorentz factors were calculated. To calculate $\theta$, the following two assumptions had to be made:

1. The $\gamma$-rays are emitted as soon as the recoiling nuclei are generated since the excited states of these nuclei often have half-lives of the order of picoseconds in which the recoiling nuclei can come out of the target but cannot travel much.

2. The motion of the recoiling nuclei is almost parallel to the beam axis as they are quite heavy and scatter less than $1\deg$.

This leads us to the conclusion that $\gamma$-ray is emitted at the central position and $\theta$ in the above equation is equal to $\theta_{TIG}$ i.e. the angle of the point of interaction in the TIGRESS detector to the beam axis. This means the Doppler correction in a particular constant angle array will be the same. Figure 4.5 shows the drift in energies for Doppler uncorrected spectra at constant angle array level. Doppler corrections improve the overall quality of peaks and help distinguish in-beam transitions from the stop lines by comparing corrected and uncorrected spectra.

### 4.1.3 Addback

A $\gamma$-ray will often scatter multiple times before being fully absorbed by the detector material. This will record a series of simultaneous energy signals across detector channels. If they are recorded as separate events, it will result in a huge background and have adverse effects on the photo-peak efficiency. However, if these events are combined by adding their energies, it will result in cleaning the background simultaneously adding to the photo-peak counts thereby increasing the photo-peak efficiency. The improvement is larger for higher energies since the probability of interaction by photoelectric effect which results in full-energy peak in a single event is low and that of scattering is high in this region.

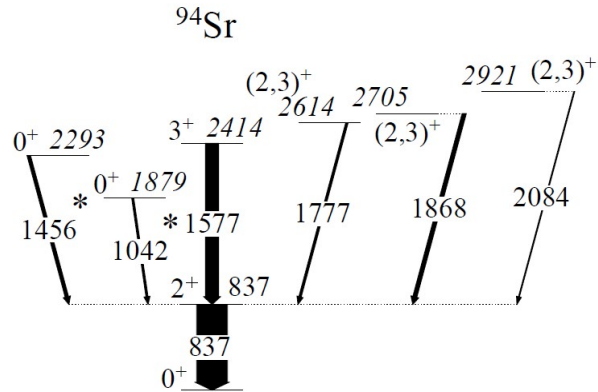

FIGURE 4.2: Level Scheme for $^{94}$Sr

The performance of this method depends on the actual algorithm used which resorts to parameters like segmentation, detector geometry of our system as well as the time difference between the signals. The hits were ordered with respect to their energies (highest to lowest) and it was considered to be the $\gamma$-ray track since the energy deposited by the $\gamma$ ray during scattering is proportional to the incident energy. This means that the position of the hit with highest energy was assumed to be the interaction point of incoming photon. This is not strictly true but is statistically most probable. This algorithm was benchmarked using calibration source data and it was found that the photo-peak efficiency of TIGRESS could be improved by up to 40% with addback.

## 4.2 SHARC

Similar to TIGRESS, analysis of SHARC was also performed. It included calibrations, gain-matching, particle identification, etc. It was done by Cruz (2017) and is not discussed here for the sake of brevity. However, it was necessary for particle-$\gamma$ coincidence to tag the $\gamma$-rays detected to a particular recoiling nuclei. The spectroscopy by and led to the construction of level scheme for $^{94}$Sr and $^{96}$Sr as shown in Figure 4.2 and Figure 4.3, respectively. Particular transitions from these level schemes were analyzed to develop the $\gamma$-ray angular distribution technique as discussed in Section 4.4.

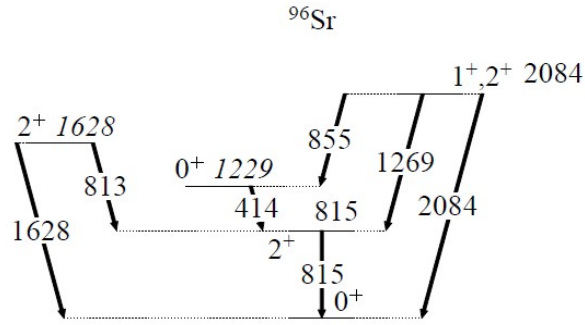

FIGURE 4.3: Level Scheme for $^{96}$Sr

## 4.3 Angular Distribution Analysis

$\gamma$-ray angular distribution was done by integrating the counts in the peak of a transition and normalizing them with the efficiencies calculated for each constant angle at that energy. Plotting the efficiency corrected counts v/s cosine of the angle of emission gives us a distribution. Fitting this distribution to Equation 1.3 gives us the value of the coefficients $a_2$ and $a_4$ which quantify the angular distribution. However, due to inverse kinematics, the system has a large center of mass frame motion which is almost relativistic. Thus, the angle values and the counts have to be corrected by transforming these values measured in laboratory frame to the values measured in the center of mass frame (Celik 2014). This is done as explained in the next subsections.

### 4.3.1 Angle Corrections

Consider a light source moving with velocity $v$ in frame $S$. In the frame $S'$ which is moving with velocity $v$ away from S, then light source is at rest. Consider a photon travelling at angle $\theta'$ as measured in $S'$. Clearly, the $x$-component of the velocity of the photon is $u'_x = c.cos\theta'$. In $S$, the $x$-component is $u_x = c.cos\theta$. From the velocity transformations:

$$u_x = \frac{u'_x + v}{1 + \frac{v.u_x}{c^2}}$$

$$c.cos\theta = \frac{c.cos\theta' + v}{1 + \frac{v.c.cos\theta}{c^2}}$$

$$cos\theta = \frac{cos\theta' + \frac{v}{c}}{1 + \frac{v.cos\theta}{c}}$$

$$cos\theta' = \frac{cos\theta - \beta}{1 - \beta.cos\theta'} \qquad (4.1)$$

where $\theta$ is the laboratory frame angle and $\theta'$ is the center of mass angle. Thus, the four angles for the constant angle array in laboratory frame were converted into center of mass frame by using Equation 4.1. The value of $\beta$ was nicely peaked around 10% with variation as little as 0.5%. Hence, it was assumed to be constant in our analysis.

### 4.3.2 Solid Angle Corrections

Due to the relativistic motion of the recoiling nuclei, it will register a different solid angle in the center of mass frame than the laboratory frame and hence, different counts in each constant angle array. However, integrating over the whole solid angle, the total counts registered in both the frames should be same since it is the same physical process viewed from two different frames of reference. We can write this as

$$W(\theta_{lab})d\Omega_{lab} = W(\theta_{cm})d\Omega_{cm}$$

That is,

$$W(\theta_{lab})sin(\theta_{lab})d\theta_{lab} = W(\theta_{cm})sin(\theta_{cm})d\theta_{cm} \qquad (4.2)$$

Using Equation 4.1 and Equation 4.2, we can write this as

$$W(\theta_{cm}) = W(\theta_{lab})\frac{(1 - \beta cos\theta_{lab})^2}{(1 - \beta^2)} \qquad (4.3)$$

It should be noted that this treatment is equivalent to solid angle corrections or efficiency corrections (as calibrations were done with the source in the center of mass frame) because the intrinsic efficiency of the detectors remains constant. The only change in the number of counts registered is due to the change in solid angles which changes the absolute efficiency. These are just multiple ways of representing the same phenomenon. Equation 4.1 and Equation 4.3 were used to plot the angular distributions in center of mass frame and find the coefficients $a'_k s$.

## 4.4 Transitions

The first step to develop this technique was to reproduce the expected angular distributions for states with assigned spin values. While choosing a transition for the same, the following points had to be considered:

1. The peak of that transition is clearly identified in the spectra and is free of any other contaminants. It should contain sufficient counts to minimize error bars. Hence, some of the strongest transitions were chosen.

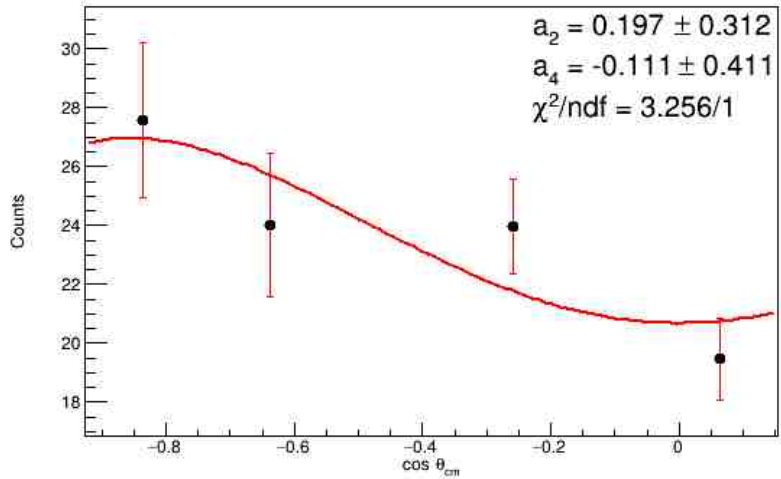

FIGURE 4.4: Angular Distribution measurements for 815 keV transition ($2_1^+ \rightarrow 0_1^+$) in $^{96}$Sr.

2. The initial state in this transition is directly populated as there are not enough statistics to perform angular correlation measurements.

3. It does not represent a stopline i.e. it shows the expected variation in the doppler uncorrected spectrum at different constant angle arrays.

4. The spins for the states involved in this transition should be accurately assigned based on other techniques like particle angular distributions and angular momentum transfer.

Of the many transitions analyzed based on the above conditions, two of the most important ones are detailed in the following subsections.

### 4.4.1   815 keV in $^{96}$Sr

This transition from the first excited state to the ground state in $^{96}$Sr is the most intense transition. Being a $2_1^+ \rightarrow 0_1^+$ transition, it is expected to be a pure $E2$ quadrupole from the selection rules. As we can see from the level scheme of $^{96}$Sr, the initial state in this transition is fed by almost all higher excited states. For performing angular distribution measurements, only the directly populated nuclei were considered and hence, strict cuts on excitation energy were made. Using the analysis techniques described above, the angular distribution measurement for this transition is as shown in Figure 4.4.

It can be seen that the distribution has a quadrupole nature but the coefficients are highly attenuated. One of the reasons for this can be poor alignment of nuclei. The alignment parameter $\sigma/J$ was back-calculated from the experimental values

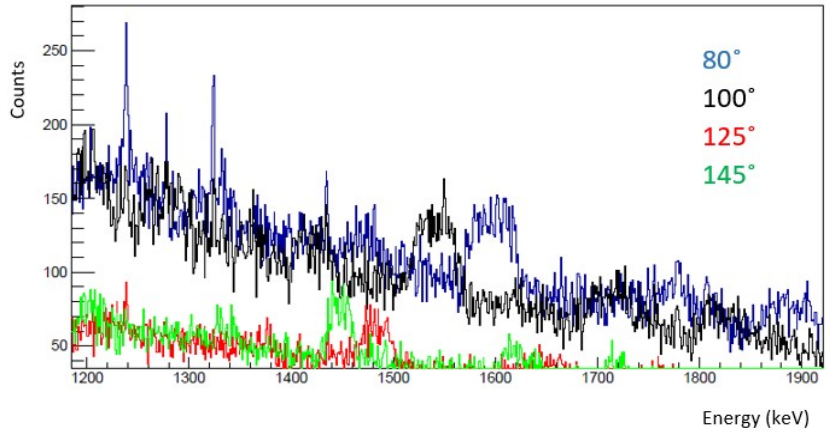

<small>FIGURE 4.5: Doppler uncorrected Spectrum at different constant array angles for 1577 keV transition ($3^- \rightarrow 2^+$) in $^{94}$Sr.</small>

for $a_2$ and $a_4$, using Equation 3.10 and the average value turned out to be $\sigma/J = 0.64 \pm 0.15$. This is in agreement with the theoretical value of $\sigma/J$ calculated in Section 3.3 thus justifying our technique.

### 4.4.2   1577 keV in $^{94}$Sr

This $3^- \rightarrow 2^+$ transition in $^{94}$Sr is interesting since the initial state is a negative parity state which is unusual in our experiments. However, this transition was detected in-beam as confirmed by the Doppler uncorrected spectrum for this transition in Figure 4.5. The parity assignment for this state is a bit vague and our analysis suggests otherwise. According to the selection rules for $\gamma$-ray transitions, it is expected to be a mixture of $E1 + M2$. Putting strict excitation energy cuts and following our analysis procedure, the angular distribution for this transition was as shown in Figure 4.6. The quadrupole nature is highly surprising.

The alignment parameter $\sigma/J$ for $J = 3$ state was theoretically calculated to be $0.4358$ in Section 3.3. Using this, the experimental mixing ratio was calculated for this transition as outlined in Section 3.2.2. The $\chi^2$ v/s $\delta$ plot for the same is shown in Figure 4.7. $\chi^2$ is minimum at $\delta = 0.91$ which shows a 91% mixing of $M2$ giving the quadrupole nature. However, according to Weiskopf's estimates, such high mixing is more probable in case of $M1 + E2$ transition which strongly suggests that the initial state should have positive parity which also checks with our experiments. Further measurements with lower uncertainties are required to confirm the same.

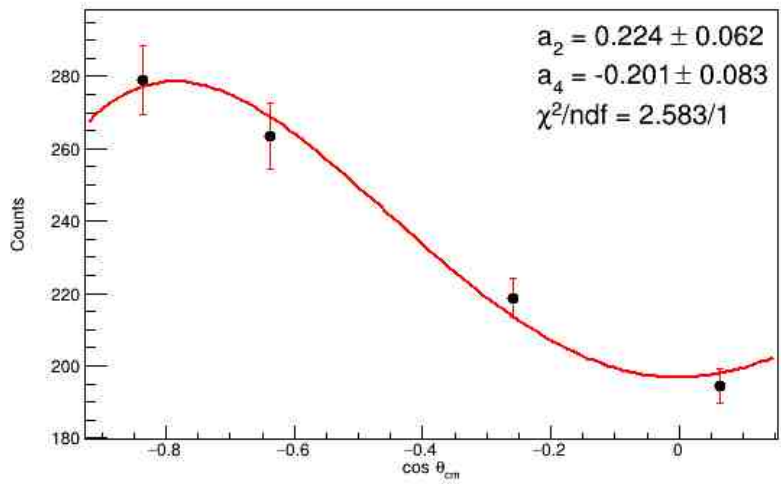

FIGURE 4.6: Angular Distribution measurements for 1577 keV transition ($3^- \rightarrow 2^+$) in $^{94}$Sr.

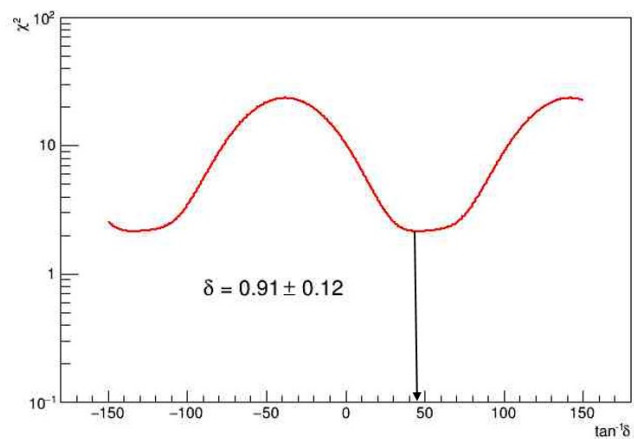

FIGURE 4.7: $\chi^2$ v/s $\delta$ plot to find the experimental mixing ratio for 1577 keV transition in $^{94}$Sr

# Chapter 5

# Conclusions and future prospects

The technique for $\gamma$-ray angular distribution measurements is developed in this work. It works well within experimental uncertainties for the states with already assigned spins. This instilled the confidence to apply this technique to other transitions of interest. Further transitions that can be analyzed include transitions like the 2084 keV in $^{96}$Sr or the 1868 keV in $^{94}$Sr. The 2084 keV transition in $^{96}$Sr is from a higher excited state to ground state ($0^+$). The spin of the higher excited state is unassigned but this transition is a pure as $J_f = 0$. Thus, we can easily assign $J_i$ based only on the nature of angular distribution.

The limiting factor for this work was the low statistics which lead to huge errors in experimentally calculated values. Higher statistics are especially required for angular distribution measurements as compared to spectroscopy since the analysis is performed at the crystal level. Another reason for the huge errors was that the coefficients $B_k$'s are fitted to the angular distribution plots from where the coefficients $a_k$'s are calculated as $a_k = \frac{B_k}{B_0}$. This leads to propagation of errors in our calculations. Higher statistics will also be required in future to perform angular correlation measurements as it was not possible in our current analysis.

Another important factor to consider for future experiments is to choose low spin beam and target nuclei wherever possible. As seen in our analysis, the high spin of deuteron target is, to an extent, responsible for the poor alignment of states in recoil nuclei as states with large m-values are populated. This, in turn, gives poor angular distributions. The TIGRESS calibrations in this work were done only with $^{152}Eu$. However, they also need to be done with appropriate sources for high energy transitions instead of extrapolating the efficiency curves from simulations since our measurements are extremely sensitive to the efficiency calibrations.

# Appendix A

# Legendre Polynomials

In principle, the angular distribution can be expressed in the form of intensity of $\gamma$-rays as a function of angle $\theta$. However, for the convenience of our analysis and calculations, these functions are expressed as a series of a set of complete functions. Legendre polynomials are an obvious choice since they are easy to work with mathematically. The coefficients in this series expansion ($a'_k s$) quantify our angular distribution measurements. Another advantage of using Legendre Polynomials is that coefficients of odd degree vanish unless the initial state is polarized. Higher order coefficients also vanish very rapidly with decreasing alignment and the highest order is restricted to $min\{2J_i, 2L_{max}\}$. So, for practical cases, we only need to work mostly with $a_2$ and $a_4$.

Legendre Polynomials are defined by the Rodrigues Formula as stated below.

$$P_l(x) = \frac{1}{2^l * l!} \frac{d^l}{dx^l}[(x^2 - 1)^l] \qquad , x = cos(\theta_\gamma) \qquad (A.1)$$

The first few even Legendre Polynomials used in our analysis can be written as follows:

$$P_0(cos(\theta)) = 1$$

$$P_2(cos(\theta)) = \frac{1}{2}(3cos^2(\theta) - 1)$$

$$P_4(cos(\theta)) = \frac{1}{8}(35cos^4(\theta) - 30cos^2(\theta) + 3)$$

These are plotted in Figure A.1

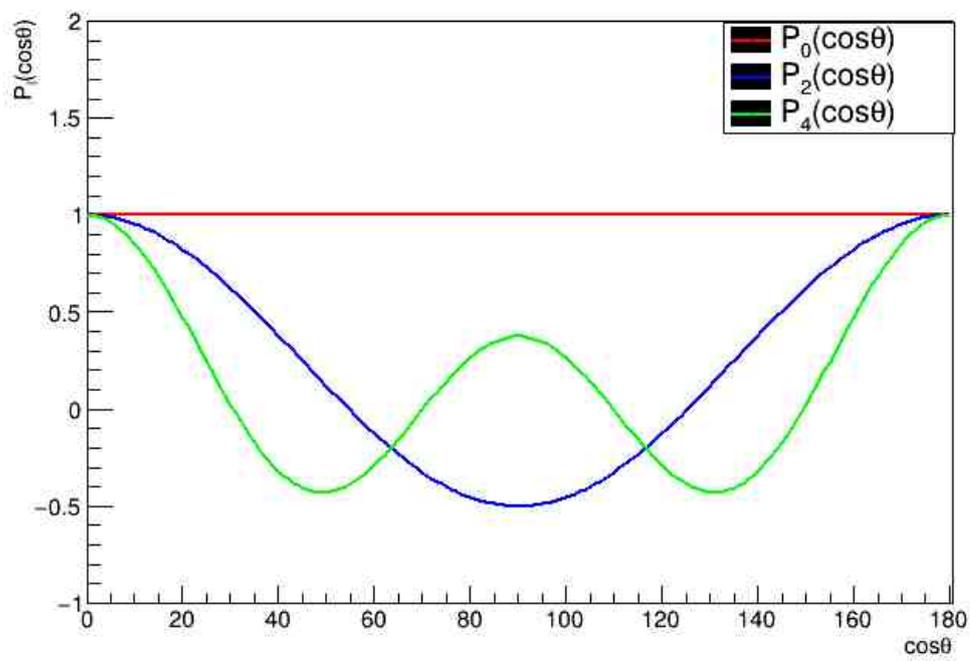

FIGURE A.1: Legendre Polynomials

# Appendix B

# Clebsch-Gordon Coefficients

In Quantum-Mechanics, coupling of two angular momenta $\vec{j_1}$ and $\vec{j_2}$ give rise to all values of angular momentum between $|j_1 - j_2|$ to $|j_1 + j_2|$. This is because the orientation of these vectors need not be collinear. This precisely means that the state $|j_1, m_1 > |j_2 m_2 >$ will not be eigenstates of the total operator $\hat{J}^2 = (\hat{J}_1 + \hat{J}_2)^2$ as eigenstates of $\hat{J}^2$ have $j$ as a good quantum number. However, the values of the projection on $\hat{z}$ axis are scalars and additive i.e. $m_1 + m_2 = m$ for any value of $\vec{j}$.

The Clebsch-Gordon coefficient $< j_1, m_1; j_2, m_2 | j, m >$ represents the amplitude of finding the $|j, m >$ state in the coupled state $|j_1, m_1 > \bigotimes |j_2, m_2 >$. More precisely, $| < j_1, m_1; j_2, m_2 | j, m > |^2$ gives the probability of finding $|j, m >$ in $|j_1, m_1 > \bigotimes |j_2, m_2 >$. However, for the Clebsch-Gordon coefficient to be non-zero, the following conditions should be met:

1. All the projection $m$-values should be less than or equal to the total angular momentum $j$-values.

2. $j_1$, $j_2$ and $j$ should follow the triangle inequality where $|j_1 - j_2| < j < |j_1 + j_2|$.

3. The projections are scalars and hence, should be additive, i.e. $m = m_1 + m_2$.

Alternatively, to obtain the state $|j, m >$, for a specified $j$, one must consider linear combinations of states $|j_1, m_1 > \bigotimes |j_2, m_2 >$ with different values of $m_1$ and $m_2$ satisfying the above conditions listed. The coefficients of this linear combination are actually the Clebsch-Gordon coefficients. Thus,

$$|j, m > = \sum_{m_1 m_2} |j_1, m_1 > |j_2, m_2 > < j_1, m_1; j_2, m_2 | j, m > \tag{B.1}$$

This immediately implies the completeness condition

$$\Sigma_{m_1 m_2} | < j_1, m_1; j_2, m_2 | j, m > |^2 = 1 \tag{B.2}$$

# Appendix C

# GRSISort

The analysis for this project was carried out using the ROOT Framework, which was adapted to suit the specific need of the experiment. The data was stored as ROOT Trees and was sorted using the GRSISort package. GRSISort was founded and developed by P.C Bender in 2013 with ongoing development by the GRIFFIN collaboration at TRIUMF. It is based on the ROOT Framework and contains additional detector classes which are designed to efficiently store the tree data and to describe in detail, the geometry of SHARC and TIGRESS detectors. The functionality of GRSISort is to produce lean trees which contain essential event information such as measured and calibrated energies, timestamps, detector element numbers, angles, etc. It contains several analysis tools which are listed below.

- TFrescoAnalysis - A program for performing Distorted Wave Born Approximation (DWBA) calculations.

- TTigressAnalysis - For analysis of Tigress Detector System including generating $\gamma$-$\gamma$ coincidence matrices.

- TSharcAnalysis - To facilitate calibration and analysis of SHARC detector system.

- TReaction - Relativistic two-body kinematics class.

- TNucleus - Nuclear data storage structure.

- TSRIM - Energy loss calculation tool using SRIM.

Apart form the above tools, additional scripts were written for grouping the TIGRESS detector system into four constant angle arrays as discussed by grouping the crystals at constant $\theta$ and summing their spectra over all $\phi$. This was done for in-beam data as well as source data for calibrations.